\documentclass[journal,12pt,onecolumn,draftclsnofoot]{IEEEtran}

\usepackage{graphicx,subfigure}
\usepackage{amsfonts,amsmath,amssymb, mathrsfs, amsthm, cite,hyperref}
\usepackage{multirow}
\usepackage{epic,eepic,eepicemu}
\usepackage{epsf}
\usepackage{epsfig}
\usepackage{graphics}
\usepackage{array} 
\usepackage{mathrsfs}
\usepackage{psfrag}
\usepackage{tikz}
\usepackage{url}
\usepackage{enumerate}
\usepackage{booktabs}
\usepackage{flushend}

\usetikzlibrary{arrows}
\usepackage{verbatim}
\newtheorem{theorem}{Theorem}

\newcommand*{\rom}[1]{\expandafter\@slowromancap\romannumeral #1@}
\makeatother

\usepackage[skip=2pt,font=footnotesize]{caption}

\DeclareMathOperator*{\argmax}{arg\,max}
\newcommand{\abs}[1]{\lvert#1\rvert}

\newcommand{\un}{\underline}
\newcommand{\be}{\begin{equation}}
\newcommand{\ee}{\end{equation}}
\newcommand{\ben}{\begin{equation*}}
\newcommand{\een}{\end{equation*}}
\newcommand{\mc}{\mathcal}

\newcommand{\e}{\epsilon}

\newcommand{\Ac}{\mathcal{A}}
\newcommand{\Cc}{\mathcal{C}}

\newcommand{\Ic}{\mathcal{I}}
\newcommand{\Tc}{\mathcal{T}}
\newcommand{\Vc}{\mathcal{V}}
\newcommand{\Xc}{\mathcal{X}}

\newcommand{\syn}{\mathsf{syn}}
\newcommand{\summ}{\mathsf{sum}}

\DeclareMathSymbol{:}{\mathbin}{operators}{"3A}

\IEEEoverridecommandlockouts

\begin{document}


\title{Coding for Segmented Edit Channels}


\author{Mahed Abroshan, Ramji Venkataramanan and Albert Guill\'en i F\`abregas\thanks{M. Abroshan and R. Venkataramanan are with the Department of Engineering, University of Cambridge, UK, (ma675@cam.ac.uk, ramji.v@eng.cam.ac.uk).}
\thanks{Albert Guill\'en i F\`abregas is with ICREA, Instituci\'o Catalana de Recerca i Estudis Avan\c{c}ats, the Department of Information and Communications Technologies, Universitat Pompeu Fabra, Barcelona, Spain, and with the Department of Engineering, University of Cambridge, UK, (guillen@ieee.org).}
\thanks{This work has been funded in part by the European Research Council under
ERC grant agreements 259663 and 725411, and by the Spanish Ministry of Economy and Competitiveness under grant TEC2016-78434-C3-1-R.}
\thanks{This paper was presented in part at the 2017 IEEE International Symposium on Information Theory, Aachen, Germany, June 2017.}
}

\maketitle

\begin{abstract}
This paper considers insertion and deletion channels with the additional assumption that the channel input sequence is implicitly divided into segments  such that at most one edit can occur within a segment. No segment markers are available in the received sequence. We propose code constructions for the segmented deletion, segmented insertion, and segmented insertion-deletion channels based on  subsets of Varshamov-Tenengolts codes chosen with pre-determined prefixes and/or suffixes.  The proposed codes, constructed for any finite alphabet, are zero-error and can be decoded segment-by-segment.  We also derive an upper bound on the rate of any zero-error code for the segmented edit channel, in terms of the segment length. This upper bound shows that  the rate scaling of the proposed codes as the segment length increases is the same as that of the maximal code.
 
\end{abstract}

\section{Introduction}

We consider the problem of constructing codes for  \emph{segmented} edit channels, where the channel input sequence is implicitly divided into disjoint segments. Each segment can undergo at most  one edit, which can be either an insertion or a deletion. There are no segment markers  in the received sequence. 

This model, introduced by Liu and Mitzenmacher \cite{liuMitz10}, is a simplified version of the general edit channel, where the insertions and deletions can be arbitrarily located in the input sequence. Constructing codes for general edit channels is well known to be challenging problem; see, e.g.,\cite{Sloane00,DaveyMackay01,ratzer05marker,abdelPFC12, cullinaKK16,brakensiekVZ16,GuruswamiW17,sala17exact}. 
The assumption of segmented edits not only simplifies the coding problem, but is also likely to hold in many edit channels that arise in practice, e.g., in data storage and in sequenced genomic data, where the number of edits is small compared to the length of the input sequence. As explained in \cite{liuMitz10}, when edits (deletions or insertions of symbols) occur due to timing mismatch between the data layout and the  data-reading mechanism, there is often a minimum gap between successive edits.  The segmented edit model includes such cases, though it also allows for nearby edits that cross a segment boundary. 
Furthermore, a complete understanding of the segmented edit model may provide insights into the open problem of constructing efficient, high-rate codes for general edit channels.  As we show in this paper,  the segmented edit assumption allows for the construction of low-complexity, zero-error codes with the optimal rate scaling for any finite alphabet.

Let us consider three examples to illustrate the model. For simplicity, we consider a binary alphabet and assume that the segment length, denoted by $b$, is  $3$ in each case. 

1) \emph{Segmented Deletion Channel}: Each segment can undergo at most one deletion; no insertions occur. Consider the following pair of input and output sequences:
\be 
X= 01{\underline{1}}\, 10\underline{0}\, 010  \  \longrightarrow  \ Y=0110010,  
\label{eq:del_example} 
\ee
with the underlined bits in $X$ being deleted by the channel to produce the  output sequence $Y$. It is easily verified that many other input sequences could have produced the same output sequence, e.g., $01{\underline{0}}\, 100\, \underline{0}10$,  $01{\underline{0}}\, 10\underline{1}\, 010$, $011\, 00\underline{0}\, 1{\underline{0}}0$ etc. The receiver has no way of distinguishing between these candidate input sequences. In particular, despite knowing the segment length and that deletions occurred, it does not know in which two segments the deletions occurred.

2) \emph{Segmented Insertion Channel}: Each segment can undergo at most one insertion; no deletions occur. The inserted bit can be placed  anywhere within the segment, including before the first bit or after the last bit of the segment. For example, consider 
\be X= 011 \, 100 \, 010 \ \longrightarrow \ Y= 011 \underline{1}  \underline{0}  100 01 \underline{1} 0, \label{eq:ins_example} \ee
with the underlined bits  in $Y$ indicating the insertions. Two inserted bits can appear between two segments whenever there is an insertion after the last bit of first segment and before the first bit of the next segment.

3)  \emph{Segmented Insertion-Deletion Channel}: This is the most general case, where a segment could undergo either an insertion or a deletion, or remain unaffected. 
For example, consider 
\be X= 01{\underline{1}} \, 100 \, 010 \ \longrightarrow \ Y= 01 \underline{0}    100 01 \underline{1} 0, \label{eq:insdel_example}\ee
with the underlined bits on the left indicating deletions, and the underlined bits on the right indicating insertions. Unlike the previous two cases, the receiver cannot even infer the exact number of edits that have occurred. In the  example above, an input sequence $9$ bits (three segments) long could result in a 10-bit output sequence in two different ways: either  via one segment with an insertion, or via two segments with insertions and the other with a deletion.

The above examples demonstrate that one cannot reduce the problem to one of correcting one edit in a $b$-bit input sequence. To see this, consider the example in \eqref{eq:del_example}, and suppose that  we used a single-deletion correcting code for each segment. Such a code would declare the first three bits of  $Y$ to be the first segment of $X$, which would result in incorrect decoding of the following segments. 

In this paper, we construct zero-error codes for each of the three segmented edit models above, for any finite alphabet of size $q \geq 2$.  Our codes can easily be constructed  even for relatively large segment sizes (several tens), and  can be decoded segment-by-segment in linear time. Moreover, the  proposed codes have rate $R$ of at least
\be 
R\geq \log_2 q -\frac{1}{b}\log_2 (b+1)-\frac{\kappa}{b} \log_2 q, \label{eq:R_LB} 
\ee 
where the constant $\kappa$ is at most $2.5$ for the segmented deletion channel, $4$ for the segmented insertion channel, and $8$ for the segmented insertion-deletion channel.  (Slightly better bounds on $\kappa$ are obtained for the binary case $q=2$.)  

We also derive an upper bound   in terms of the segment length $b$ on the maximum rate of any code for the segmented edit channel. This upper bound (Theorem \ref{thm:rate_UB}) shows that the rate $R$ of any zero-error code with code length $n$  satisfies
\be
R \leq  \log_2 q-  \frac{1}{b} \log_2 b - \frac{1}{b} \log_2 (q-1)+ \frac{1}{b} +\frac{\log_2 (2q)}{n}+ O\left( \frac{\ln b}{b^{4/3}} \right). 
\label{eq:R_UB}
\ee
Comparing \eqref{eq:R_LB} and \eqref{eq:R_UB}, we see that the rate scaling for the proposed codes is the same as that of the maximal code with the rate penalty being $O(1/b)$. 

The starting point for our code constructions is the family of Varshamov-Tenengolts (VT) codes \cite{VT65,Sloane00, Tenengolts84}. Each code in this family is a single-edit correcting code.  
In our constructions, the codewords in each segment are drawn from subsets of VT codes satisfying certain prefix/suffix conditions, which are carefully chosen to enable fast segment-by-segment VT decoding.

\subsection{Comparison with previous work} 
The segmented edit assumption places a restriction on the kinds of edit patterns that can be introduced in the input sequence. Other models with restrictions on edit patterns include the forbidden symbol model considered in \cite{kulkarni14forbidden}. 

We now highlight some similarities and differences from the codes proposed by Liu and Mitzenmacher in  \cite{liuMitz10} for the binary segmented deletion and segmented insertion channels.

\emph{Code construction}: The code in \cite{liuMitz10} is a binary segment-by-segment code specified via sufficient conditions \cite[Theorems 2.1, 2.2]{liuMitz10} that ensure that as decoding proceeds, there are at most two choices for the starting position of the next undecoded segment. Finding the maximal code that satisfies these conditions corresponds to an independent set problem, which is challenging for large $b$.  The maximal code satisfying these conditions was reported in \cite{liuMitz10} for $b=8,9$. For larger $b$, a greedy algorithm was used to find a set of codewords satisfying the conditions. It was also suggested that one could restrict the code to a subset of VT codes that satisfy the sufficient conditions. 

In comparison, our codes are  directly defined as subsets of VT codes that satisfy certain simple prefix/suffix conditions; these conditions are different from those in \cite{liuMitz10}. Our conditions ensure that upon decoding each segment, there is no ambiguity in the starting position of the next segment. These subsets of VT codes are relatively simple to enumerate, so it is possible to find the largest code satisfying our conditions for $b$ of the order of several tens. Table \ref{tab:code_rate} lists the number of codewords per segment for the three segmented edit channels for $q=2$ and lengths up to $b=24$. For the segmented deletion and segmented insertion-deletion channels, another difference from  the code in \cite{liuMitz10}  is that our codebook for each segment is chosen based on the final bit of the previous segment.

\emph{Rate}:  The VT subsets and sufficient conditions we define allow us to obtain a lower bound of the form \eqref{eq:R_LB} on the rate of our code for any segment length $b$.   Though the \emph{maximal} codes satisfying the Liu-Mitzenmacher conditions have rate very close to the largest possible  with segment-by-segment decoding, finding the maximal code satisfying these conditions is computationally hard,  so one  has to resort to greedy algorithms to construct codes for larger $b$.
This is reflected in the rate comparison:  for $b=8,9$, the optimal Liu-Mitzenmacher code for segmented deletions is larger than our code (12,20  vs. 8,13 codewords). However for $b=16$, the code obtained in \cite{liuMitz10} using a greedy algorithm has  652 codewords, whereas   our code has $964$ codewords, as shown in Table \ref{tab:code_rate}.  For large $b$, our codes are nearly optimal since the rate penalty decays as $\kappa/b$.

For the segmented insertion channel,  it is shown in Sec. \ref{subsec:LM}  that our code construction  satisfies the sufficient conditions specified \cite{liuMitz10}. The lower bound on the rate of our code affirmatively answers the conjecture in \cite{liuMitz10} that the rates of the maximal codes satisfying the sufficient conditions increases with $b$. 


\emph{Encoding and decoding complexity}: Being subsets of VT codes, our codes can also be efficiently encoded even for large segment sizes $b$,  without the need for look-up tables \cite{AbdelFer98,VTEncodingAVG17}. As segment-by-segment decoding is enforced by design, the decoding complexity grows linearly with the number of segments for both our codes and those in \cite{liuMitz10}.  Within each segment, the decoding complexity of our code is also linear in $b$, since VT codes can be decoded with linear complexity \cite{Sloane00}. In general, for each segment, the maximal Liu-Mitzenmacher codes have to be decoded via look-up tables, in which case the complexity is exponential in $b$.  Using subsets of VT codes was suggested in \cite{liuMitz10} as a way to reduce the decoding complexity.

Finally, we remark that codes proposed in this paper  are the first for the binary segmented insertion-deletion model, and for all the non-binary segmented  edit models.

\begin{table}[t!]
\centering
\caption{Number of codewords per segment of the proposed codes. Lower bounds computed from \eqref{eq:Rdel_LB}, \eqref{eq:rate_ins},  and \eqref{eq:rate_indel} are given  in brackets.\vspace{1mm}}
\begin{tabular}{cccc} 
\toprule
 $b$ & Deletion & Insertion & Insertion-Deletion\\
\midrule   
		  8    &  8 (8)  			 & 6 (6)  			 &  1 (1)		   \\
          9    &  13 (13)     		 & 10  (10)   		 &  2 (1)		   \\
          10   &  24 (24)    		 & 18  (18)    		 &  2 (1)		   \\
          11   &  44 (43)     		 & 33  (32)     	 &  2 (2)		   \\
          12   &  79  (79)     		 & 60  (59)    		 &  4 (3)		   \\
          13   &  147  (147)    	 & 111 (110)   		 &  6 (5)		   \\
          14   &  276  (274)     	 & 208 (205)    	 &  12 (9)		   \\
          15   &  512   (512)     	 & 384 (384)     	 &  16 (16)	   	   \\
          16   &  964  (964)     	 & 724 (723)         &  34 (31)		   \\
          17   &  1,824  (1,821)     & 1,368 (1,366)     &  59 (57)		   \\
          18   &  3,450 (3,450)      & 2,588 (2,587)     &  114 (108)	   \\
          19   &  6,554  (6,554)     & 4,916 (4,916)     &  206 (205)	   \\
          20   &  12,490 (12,484)    & 9,369 (9,363)     &  399 (391)	   \\
          21   &  23,832 (23,832)    & 17,847 (17,874)   &  746 (745)	   \\
          22   &  45,591 (45,591)    & 34,194 (34,193)   &  1,435 (1,425)  \\
          23   &  87,392  (87,382)   & 65,544 (65,536)   &  2,736 (2,731)  \\
          24   &  167,773 (167,773)  & 125,831 (125,830) &  5,257 (5,243)  \\
           \bottomrule 
\end{tabular}
\label{tab:code_rate}
\end{table}
   
 \subsection{Organization of the paper}
 The remainder of the paper is organized as follows.  In Section \ref{sec:ch_model}, we formally define the channel model, and review binary and non-binary VT codes.  In Section \ref{sec:rate_UB}, we derive an upper bound on the rate of any code for a segmented edit channel, in terms of the segment length. In Sections \ref{sec:del}, \ref{sec:ins}, and \ref{sec:indel}, we present our code constructions for the segmented deletion channel, segmented insertion channel, and the segmented insertion-deletion channel, respectively. For each model, we first treat the binary case to highlight the key ideas, and then extend the construction to general non-binary alphabets. 

 
\section{Channel Model and Preliminaries} \label{sec:ch_model}
The channel input sequence is denoted by $X=x_1x_2\cdots x_n$, with $x_i\in\Xc$ for $i=1,\dotsc,n$, where $\Xc=\{0,\dotsc,q-1\}$ is the input alphabet, with $q \geq 2$. 
The channel input sequence is divided into $k$ segments of $b$  symbols each. We denote the subsequence of $X$, from index $i$ to index $j$, with $i<j$ by $X(i : j)=x_ix_{i+1}\cdots x_j$.
The $i$-th segment of $X$ is denoted by $S_i = s_{i,1}\cdots s_{i,b} = X\big (b(i-1)+1:bi\big) $ for $i=1,\dotsc,k$.

In the segmented deletion channel, the channel output $Y= Y(1 : m)= y_1\cdots y_m$, with $m\leq n$ is obtained by deleting at most one symbol in each segment, i.e., at most one symbol in $S_i$, $i=1,\dotsc,k$, is deleted. Similarly, in the segmented insertion channel, the channel output $Y=y_1\dotsc y_m$, with $m\geq n$ is obtained by inserting at most one symbol per segment. In the segmented insertion-deletion channel, the channel output is such that each segment $S_i$, $i=1,\dotsc,k$ undergoes at most one edit. In all cases, we assume that the decoder knows $k$ and $b$, but not the segment boundaries. 

We consider coded communication using a code $\Cc = \{X^{(1)},\dotsc,X^{(M)}\}\subseteq \Xc^n$ of length $n$, $M$ codewords and rate $R=\frac{1}{n}\log_2 M$. We consider segment-by-segment coding, where $M_s$ is the number of codewords per segment. The overall code of length $n=kb$ has $(M_s)^k$ codewords, and rate 
\begin{align}
R &= \frac{1}{n}\log_2 (M_s)^k \\
&= \frac{1}{b}\log_2 M_s.
\end{align} 

The decoder produces an estimate $\hat X$ of the transmitted sequence. We denote the corresponding segment estimates by $\hat S_i= \hat s_{i,1}\cdots \hat s_{i,b}$, for $i=1,\dotsc,k$. Thus $\hat{X} = (\hat S_1, \ldots, \hat S_k)$. We consider zero-error codes that always ensure the recoverability of the transmitted sequence, i.e., codes for which $\hat X = X$.

\subsection{Binary VT codes}
First consider the case where $q=2$, i.e., $\mc{X}=\{ 0,1\}$.
Suppose that $k=1$,  and thus $n=b$, i.e., there is at most one  edit in the entire  binary sequence.  For this model, one can use binary VT codes which are zero-error 
single-edit correcting codes \cite{VT65, Sloane00}, i.e.,  when the transmitted codeword suffers a single insertion or a deletion, the decoder always corrects the edit. 
Moreover, the complexity of the VT decoding algorithm is  linear in the code length $b$. The details of the decoding algorithm can be found in \cite{Sloane00} for the case of a single deletion; the decoding algorithm to correct from a single insertion can be found in \cite[Sec. II]{Tenengolts84}.

The VT syndrome of a binary sequence $S =s_1 \ldots s_b$  is defined as 
\begin{equation}
\syn(S)=\sum_{j=1}^b j \,s_j \  \  (\text{mod}  (b+1)).
\end{equation}
For positive integers $b$ and $0\leq a\leq b$, we define the VT code of length $b$ and syndrome $a$, denoted by 
\be
\Vc\Tc_a(b) = \big\{S \in \{0,1\}^b: \syn(S)=a\big\} 
\ee
i.e., the set of sequences $S$ of length $b$ that satisfy $\syn(S)=a$. For example, 
\be
\Vc\Tc_1(3) =  \Bigl\{s_1s_2s_3:\sum_{j=1}^3 j\,s_j = 1 \text{ mod 4} \Bigr\}=\{100, 011  \}.
\ee 
The $b+1$ sets $\Vc\Tc_a(b) \subset \{0,1\}^b$, for $0\leq a\leq b$, partition the set of all sequences of length $b$.  Each of these sets $\Vc\Tc_a(b)$ is a single-edit correcting code. In particular, if $S, S' \in \Vc\Tc_a(b)$, then 
\be
\mc{D}_1(S) \cap \mc{D}_1(S') = \emptyset,  \ \text{ and } \  \mc{I}_1(S) \cap \mc{I}_1(S') = \emptyset,
\label{eq:D1I1_prop}
\ee
where $\mc{D}_1(S)$ denotes the set of subsequences  obtained by deleting one bit from $S$, and $\mc{I}_1(S)$  is the set of supersequences obtained by inserting one bit in $S$.

 For   $0 \leq a \leq b$,  the cardinalities of these sets satisfy \cite[Corollary 2.3]{Sloane00}
\be
 \abs{\Vc\Tc_0(b)} \geq \abs{ \Vc\Tc_a(b)} \geq \abs{ \Vc\Tc_1(b)}. 
 \ee
The largest of the sets $\Vc\Tc_a(b), \ 0 \leq a \leq b$, will have at least $\frac{2^b}{b+1}$ sequences out of the $2^b$ possible. This induces a rate $R \geq 1 - \frac{1}{b}\log_2(b+1)$ for the largest of these codes. 
The code  $\Vc\Tc_0(b)$ has been shown to be maximal  for single edit correction for $b\leq 8$, and has been conjectured to be maximal for arbitrary $b$ \cite{Sloane00}. 

\subsection{Non-binary VT codes}
Here we consider the case where $\mc{X} = \{0, \ldots, q-1 \}$, with $q >2$. Again, suppose that $k=1$ and thus $n=b$, i.e.,  there is at most one  edit in the  sequence. For this model, one can use $q$-ary VT codes, introduced by Tenengolts \cite{Tenengolts84}. These are zero-error single-edit correcting codes, analogous to the binary VT codes.  We briefly describe the code construction below. 

For each non-binary sequence $S$, define a length $(b-1)$ auxiliary binary sequence $A_S=\alpha_2,\dotsc,\alpha_b$   as follows. For $2 \leq i \leq b$,
\be \alpha_i =
  \begin{cases}
    1  & \quad \text{if } s_i\geq s_{i-1}\\
    0  & \quad \text{if } s_i< s_{i-1} \\
  \end{cases}
\ee
We also define the modular sum as
\be
\summ(X)=\sum_{i=1}^b s_i \quad (\text{mod } q).
\ee
The $q$-ary VT code with length $b$ and  parameters $(a,c)$ is defined as \cite{Tenengolts84}
\be
\Vc\Tc_{a,c}(b) = \big\{S \in \Xc^b: \syn(A_S)=a,\, \summ(S)=c\big\},
\ee
for $0\leq a\leq b-1$ and $c\in \Xc$. Similarly to the binary case, the sets $\Vc\Tc_{a,c}(b)$  for $0\leq a\leq b-1$ and $c\in \Xc$ partition the space $\Xc^b$ of all $q$-ary sequences of length $b$. Clearly, the largest codebook has at least $\frac{q^b}{qb}$ codewords which implies the following rate lower bound for the largest VT code among all choices of $(a.c)$:
\begin{equation}
R\geq \log_2 q-\frac{1}{b}\log_2 b-\frac{1}{b}\log_2 q.
\end{equation}

The complexity of the decoding algorithm for  $q$-ary VT codes is linear in the code length $b$. The details of the decoder can be found in \cite[Sec. II]{Tenengolts84}.

\section{Upper Bound on Rate} \label{sec:rate_UB}

In this section, we derive an upper bound on the rate of any code for $q$-ary segmented edit channels, for $q\geq 2$. The upper bound is valid for all zero-error codes, including those that cannot be decoded segment-by-segment. 
\begin{theorem}
For each of the three segmented edit models, with segment length $b$, the rate $R$  of any zero-error code with code length $n=kb$ satisfies
\be R \leq  \log_2 q-  \frac{1}{b}\log_2 b -  \frac{1}{b}\log_2 (q-1) +  \frac{1}{b} +\frac{\log_2 (2q)}{kb}+ O\left( \frac{\ln b}{b^{4/3}} \right). \ee  
\label{thm:rate_UB}
\end{theorem}
\emph{Remarks}:
\begin{enumerate}
\item In the theorem, the alphabet size $q$ is held fixed as the segment size $b$ grows. The number of segments per codeword, $k$, is arbitrary, and need not grow with $b$.

\item  The theorem is obtained via  non-asymptotic bounds on the size and the rate of any zero-error code. These bounds, given in \eqref{eq:Mbound}--\eqref{eq:logT2}, may be of independent interest. 

\item The dominant terms in the upper bound may be interpreted as follows for the case of the segmented deletion channel. For a noiseless $q$-ary input channel the rate is $\log_2 q$ bits/transmission. The $\log_2 b /b$ term corresponds to a penalty required to convey the run  in which  the deletion occurred in each segment. The $\log_2(q-1)/b$ term is  a penalty required to convey the value of the deleted symbol. 
\end{enumerate}
\begin{IEEEproof}[Proof of Theorem \ref{thm:rate_UB}]
We give the proof for the segmented deletion model with segment length $b$. The argument for the segmented insertion model is similar.

The proof technique is similar to that used by Tenengolts in \cite[Theorem 2]{Tenengolts84}. The high-level idea is the following. The codewords are split into two groups: the first group contains the codewords in which a large majority of segments have at least $b \frac{(q-1)}{q} - O(b^{2/3})$ runs. The other group contains the remaining codewords. As $b$ grows larger, the fraction of length $b$ sequences with close to $b \frac{(q-1)}{q}$ runs  (the `typical' value) approaches 1. So we carefully bound the number of codewords in the first group, while the number of codewords in the second group can be bounded by a direct counting argument.

Consider a code $\Cc$ of length $n=kb$, i.e., each codeword  has $k$ segments of length $b$. Let $M = \abs{C} =2^{nR}$ denote the size of the code. For integers $r \geq 0$ and $0 \leq l \leq k$,  define 
$\mc{M}(r,l) \subset \Cc$ as the set of the codewords that have exactly $l$ segments with more than $r$ runs. Let $M(r,l) = \abs{\mc{M}(r,l)}$.  Note that for any $r \geq 0$, we have \be \sum_{l=0}^k M(r,l)=M.  \label{eq:Mrl_sum} \ee

For any $l \leq k$ and a codeword $x \in \mc{M}(r,l)$, let $\rho_l(x)$ denote the number of distinct sequences of length $(n-l)$ by deleting exactly $l$ symbols from $x$ (following the segmented assumption). We then have
\be (r-1)^l \leq \rho_l(x). \label{eq:rk_bnd} \ee
To show \eqref{eq:rk_bnd}, we only need to consider  $r \geq 3$ as the inequality is trivial for $r \leq 2$.  Considering the $l$ segments that each have at least $(r+1)$ runs,  there are at least 
$(r-1)^l$ ways of choosing one run from each segment so that the $l$ chosen runs are non-adjacent.  For each such choice of $l$ non-adjacent runs, we get a distinct subsequence of length $(n-l)$ by deleting one symbol from each run. This proves \eqref{eq:rk_bnd}.

Since $\Cc$ is a zero-error code, for two distinct codewords $x_1, x_2 \in \mathcal{M}(r,l)$, the set of length $(n-l)$ sequences obtained via  $l$ deletions (in a segmented manner) from $x_1$ must be distinct from the corresponding set for codeword $x_2$. We therefore have
\begin{align}
q^{n-l} &\geq \sum_{x\in \mathcal{M}(r,l)}  \rho_l(x) \\
&\stackrel{(a)}{\geq} \sum_{x\in \mathcal{M}(r,l)}(r-1)^l \\
&= M(r,l)(r-1)^l,
\end{align}
where $(a)$ is obtained from  \eqref{eq:rk_bnd}. We therefore obtain
\be
M(r,l) \leq \frac{q^{n-l}}{(r-1)^l}.
\label{eq:mrk_bnd}
\ee
Fix $\alpha \in (0,1)$. Summing \eqref{eq:mrk_bnd} over $ \alpha k \leq l \leq k$, we obtain
\begin{align}
\sum_{l \geq \alpha k} M(r,l) &\leq  \sum_{l \geq \alpha k}   \frac{q^{n-l} }{(r-1)^l} \\
&\leq \frac{2 q^{n- \alpha k}}{(r-1)^{\alpha k}}.
\end{align}
Now choose
\be r= \frac{(q-1)}{q}b- \sqrt{ \frac{2 \kappa (q-1) b \ln b}{q}},  \label{eq:rdef} \ee
 where $\kappa >\frac{\log (2q)}{\log b}$ will be specified later.  Using this $r$ in \eqref{eq:mrk_bnd}, and noting that $n=kb$, we have
\begin{align} 
\sum_{l \geq \alpha k} M(r,l)  &\leq \frac{2 q^{kb -\alpha k }}{(r-1)^{\alpha k}}\\
&= \frac{2 q^{kb}}{(b(q-1))^{\alpha k} \left( 1 -  \sqrt{ \frac{2  \kappa q \ln b}{(q-1)b}} - \frac{q}{(q-1)b} \right)^{\alpha k}}.\label{eq:l_upper}
\end{align}
For $l < \alpha k$, we use the  looser bound
\begin{equation}
M(r,l)\leq  {k \choose k-l} \left[ q \sum_{t=0}^{r-1} (q-1)^{t}{b-1 \choose t} \right]^{k-l}q^{bl}, 
\label{eq:qMrl}
\end{equation}
which is obtained as follows.  We first choose the $(k-l)$ segments with at most $r$ runs. Then, a segment with $t$ runs is determined by the choice of the first symbol, and the starting positions and values of the next $(t-1)$ runs. There are $q$ choices for the first symbol, ${b-1 \choose t-1}$ choices for the starting position of the next $(t-1)$ runs, and $(q-1)^{t-1}$ choices for the values of these runs. Therefore, the number of possible  length $b$ sequences with at most $r$ runs is $ q \sum_{t=1}^{r} {b-1 \choose t-1} (q-1)^{t-1}=q \sum_{t=0}^{r-1} {b-1 \choose t} (q-1)^t$.
We then obtain \eqref{eq:qMrl}  by noting that: i) there are $(k-l)$ segments with at most $r$ runs, and ii) there are at most $q^{bl}$ choices for the remaining $l$ segments.
 
We write the  right hand side of  \eqref{eq:qMrl} as 
  \begin{align}
 {k \choose k-l}  \left[ q \sum_{t=0}^{r-1} (q-1)^{t}{b-1 \choose t} \right]^{k-l}q^{bl}  
 &  =  {k \choose k-l} \left[ q^{b+1} \sum_{t=0}^{r-1} \left(1-\frac{1}{q}\right)^{t} \left( \frac{1}{q}\right)^{b-t}  {b-1 \choose t} \right]^{k-l}q^{bl}  \\
 &  \leq   2^k q^{bk+k-l} \left[ \sum_{t=0}^{r-1} \left(1-\frac{1}{q}\right)^{t} \left( \frac{1}{q} \right)^{b-t} {b-1 \choose t} \ \right]^{k-l}.
 \label{eq:RHS}
 \end{align}
 
 It is shown in Appendix \ref{app:binom_tail} that 
 \be
 \sum_{t=0}^{r-1} \left(1-\frac{1}{q}\right)^{t} \left( \frac{1}{q} \right)^{b-t} {b-1 \choose t}  \leq \frac{1}{b^\kappa}.
 \label{eq:binom_tail}
 \ee
Using \eqref{eq:binom_tail} to bound \eqref{eq:RHS}, and then substituting in \eqref{eq:qMrl}, we obtain
\be
M(r,l) \leq \frac{2^k q^{bk +k -l}}{b^{\kappa (k-l)}}.
\ee
Summing over $0 \leq l < \alpha k$ and considering $\kappa >\frac{\log (2q)}{\log b}$, we obtain
\begin{align}
\sum_{l < \alpha k} M(r,l) &\leq \frac{2^k q^{(b+1)k}}{b^{\kappa k}}\sum_{l < \alpha k}\left(\frac{b^\kappa}{q}\right)^l \\
&\leq \frac{2^k q^{(b + 1-\alpha)k +1}}{b^{\kappa(1-\alpha)k}}.\label{eq:l_lower}
\end{align}
Combining the bounds in \eqref{eq:l_upper} and \eqref{eq:l_lower}, we have 
\begin{align}
M &= \sum_{l = 0}^k M(r,l) \\
&\leq \frac{2 q^{kb}}{(b(q-1))^{\alpha k} \left( 1 -  \sqrt{ \frac{2  \kappa q \ln b}{(q-1)b}} - \frac{q}{(q-1)b} \right)^{\alpha k}} 
+ \frac{2^k q^{(b + 1-\alpha)k +1 }}{b^{\kappa(1-\alpha)k}} \\
&\leq 2 \max \{ T_1, T_2 \} \label{eq:Mbound}
\end{align}
where
\be
T_1=  \frac{2 q^{kb}}{(b(q-1))^{\alpha k} \left( 1 -  \sqrt{ \frac{2  \kappa q \ln b}{(q-1)b}} - \frac{q}{(q-1)b} \right)^{\alpha k}}, \qquad 
T_2= \frac{2^k q^{(b + 1-\alpha)k +1}}{b^{\kappa(1-\alpha)k}}.
\label{eq:T1T2def}
\ee
Therefore the rate can be bounded as 
\be
R = \frac{\log M}{kb} \leq \frac{1}{kb} + \max \left \{ \frac{\log T_1}{kb}, \ \frac{\log T_2}{kb} \right \}.
\ee
From \eqref{eq:T1T2def}, we have
\begin{align} \label{eq:logT1a}
\frac{\log T_1}{kb} & \leq  \log_2 q  - \frac{\alpha  \log_2 (b(q-1))}{b} - \frac{\alpha }{b} \log_2\left( 1-   \sqrt{ \frac{2 \kappa q \ln b}{(q-1)b}} - \frac{q}{(q-1)b} \right)   + \frac{1}{kb}, \\
\frac{\log T_2}{kb}& \leq   \log_2 q  - \frac{ \kappa (1-\alpha)  \log_2 b}{b}  + \frac{(1-\alpha) \log_2 q}{b} + \frac{1}{b} +  \frac{\log_2 q}{kb}.
\label{eq:logT2}
\end{align}
Now choose $\alpha$ and $\kappa$ as follows:
\begin{align} \label{eq:alpha_choice}
\alpha &= 1- \frac{1}{\sqrt[3]{b}},\\
\kappa &= \frac{\alpha}{1-\alpha} \frac{\log_2 (b(q-1))}{\log_2 b} \\
&= \left(\sqrt[3]{b} -1 \right)  \frac{\log_2 (b(q-1))}{\log_2 b}.
\label{eq:kappa_choice}
\end{align}
Note that we have $\alpha \to 1$ and $\frac{2 \kappa q \ln b}{(q-1)b} \to 0$ as $b \to \infty$.
Using the fact that $\ln(1/(1-x)) \leq 2 x$ for $x \in (0,1/2]$ in \eqref{eq:logT1a},  we have the following bound on $T_1$ for sufficiently large $b$:
\be
\begin{split}
& \frac{\log T_1}{kb} \leq  \log_2 q  - \frac{\alpha  \log_2 (b(q-1))}{b} + \frac{1}{kb} + \frac{2\alpha }{b \ln 2} \left(  \sqrt{ \frac{2 \kappa q \ln b}{(q-1)b}} + \frac{q}{(q-1)b} \right)    \\
& = \log_2 q  - \frac{\log_2 (b(q-1))}{b}+ \frac{\log_2 (b(q-1))}{b^{4/3}}+  \frac{1}{kb} + \frac{2\alpha }{b \ln 2} \left(  \sqrt{ \frac{2 \kappa q \ln b}{(q-1)b}} + \frac{q}{(q-1)b} \right).
\end{split}
\label{eq:logT1}
\ee
Also substituting the values of $\alpha, \kappa$ from \eqref{eq:alpha_choice} and \eqref{eq:kappa_choice} in \eqref{eq:logT2}, we have
\be
\frac{\log T_2}{kb} \leq   \log_2 q  - \frac{\log_2 (b(q-1))}{b}+ \frac{1}{b} + \frac{\log_2 (b(q-1))}{b^{4/3}}  + \frac{\log_2 q}{b^{4/3}} +  \frac{\log_2 q}{kb}.
\label{eq:logT2:2}
\ee
Finally, substituting the values of $\alpha, \kappa$ into the last term in \eqref{eq:logT1},  it can be seen that this term is $O(\sqrt{\ln b}/b^{4/3})$, which  yields the desired result. 
 \end{IEEEproof}

\section{Segmented Deletion Correcting Codes} \label{sec:del}

In this section, we show how to construct a segment-by-segment zero-error code for the segmented deletion channel. For simplicity, we first introduce binary codes and explain the binary decoder. We then highlight the differences in the non-binary case.

If the decoder knew the segment boundaries, then simply using a VT  code for each segment would suffice. Since the segment boundaries are not known, recall from the example in \eqref{eq:del_example} that this approach is inadequate if segment-by-segment decoding is to be used. Our construction chooses a subset of a VT code for each segment, with prefixes determined by the last symbol of the previous segment.

\subsection{Binary Code Construction} \label{subsec:binary_deletion}

For $0 \leq a \leq b$, define the following sets.
\begin{align}
\begin{split}
&\Ac^0_a \triangleq \big\{S \in \{0,1\}^b : \syn(S)=a, \ s_1s_2=00 \big\}, \\
&\Ac^1_a \triangleq \big\{S \in \{0,1\}^b : \syn(S)=a, \ s_1s_2=11\big\}.
\end{split}
\end{align}
For $c\in \{ 0,1 \}$, the set $\Ac^c_a\subseteq \Vc\Tc_a(b)$ is the set of VT codewords that start with prefix $cc$. We now choose the sets with the largest number of codewords, i.e., we choose $\Ac^0_{a_0}$ and $\Ac^1_{a_1}$ where we define 
\be
a_0 = \argmax_{0\leq a \leq b} |\Ac^0_a|, \quad a_1 = \argmax_{0\leq a \leq b} |\Ac^1_a|.
\ee
By defining $M_s=\min\{|\Ac^0_{a_0}|,|\Ac^1_{a_1}|\}$, we can now construct  $\Ac^0\subseteq \Ac^0_{a_0}$ by choosing any $M_s$ sequences from $\Ac^0_{a_0}$; similarly construct $\Ac^1\subseteq \Ac^1_{a_1}$ by choosing any $M_s$ sequences from $\Ac^1_{a_1}$.  The sets $\Ac^0$ and $\Ac^1$ are subsets of the VT codes $\Vc\Tc_{a_0}(b)$ and $\Vc\Tc_{a_1}(b)$, containing sequences starting with $00$ and $11$, respectively.

Finally, the overall code of length $n=kb$  is constructed  by choosing a codeword for each segment from either $\Ac^0$ or $\Ac^1$. The codeword for the first segment is chosen from $\Ac^0$. The codeword for segment $i =2,\dotsc,k$ is chosen as follows: if the last code bit in segment $(i-1)$ equals $0$, then the codeword for segment $i$ is chosen from $\Ac^{1}$; otherwise it  is chosen from $\Ac^{0}$.  

\subsection{Rate}
The rate of the above codes can be bounded from below as 
\be
R \geq 1-\frac{1}{b}\log_2 (b+1)-\frac{2}{b}.
\label{eq:Rdel_LB}
\ee
Indeed, there are $2^{b-2}$ binary sequences of length $b$ whose first two bits equal $0$. Each of these sequences belongs to exactly one of the sets $\Ac^0_0, \ldots, \Ac^0_b$. Therefore, the largest among these $(b+1)$ sets will contain at least $2^{b-2}/(b+1)$ sequences and thus, 
\be
|\Ac^0_{a_0}|  \geq \frac{2^{b-2}}{b+1}.
\ee 
A similar  argument gives the same lower bound for $|\Ac^1_{a_1}|$, hence 
\be
M_s  \geq \frac{2^{b-2}}{b+1}.
\ee
Taking logarithms gives \eqref{eq:Rdel_LB}.

From \eqref{eq:Rdel_LB},  we see that the rate penalty with respect to VT codes is at most $\frac{2}{b}$ due
to the prefix of length 2. As an example, for $b=16$ our code has $964$ codewords, while the greedy algorithm described in \cite{liuMitz10}, gives $740$; this is reduced to $652$ when the search is restricted to VT codes. More examples are reported in Table \ref{tab:code_rate}.

\subsection{Decoding}

Thanks to the segment-by-segment code construction, decoding will also proceed segment by segment. 
Decoding proceeds in the following simple steps.

In order to decode segment $i$, for $i=1,\dotsc,k$, assume that the first $i-1$ segments have been decoded correctly. Thus the decoder knows the correct starting position of segment $i$ in $Y$; we denote it by $p_i+1$.

By examining the last bit of segment $(i-1)$,  the decoder learns  the correct syndrome for the codeword in segment $i$, i.e., either $a_0$ or $a_1$; recall that segment 1 was drawn from $\Ac^0$. Without loss of generality, assume it is $a_0$; the decoding for $a_1$ is identical.

\begin{enumerate}
\item The decoder computes the VT syndrome
\be
\hat a = \syn\bigl(Y(p_i+1: p_i+b)\bigr)
\label{eq:bin_dec_del}
\ee
and compares it to the correct syndrome (assumed to be $a_0$). There are two possibilities:
\begin{enumerate}
\item $\underline{\hat a = a_0}$: The decoder concludes that there is no deletion in segment $i$. This is because  if there was a deletion in segment $i$, then $Y(p_i+1: p_i+b)$ cannot have VT syndrome $a_0$ unless $Y(p_i+1: p_i+b) = S_i$ --- indeed,   if $Y(p_i+1: p_i+b) \neq S_i$, then both  these length $b$ sequences would have syndrome $a_0$ and  $Y(p_i+1: p_i+b-1)$ as a subsequence, contradicting the property of VT codes in \eqref{eq:D1I1_prop}.

In this case, the decoder  outputs $\hat S_i=Y\bigl(p_i+1:p_i+b\bigr)$. The starting position of the next segment in $Y$ is $p_i+b+1$.

\item $\underline{\hat a \neq a_0}$: The decoder knows there is a deletion in segment $i$ and feeds $Y\bigl(p_i+1:p_i +b-1\bigr)$ to the VT decoder to recover the codeword. The output of the VT decoder is the decoded segment $\hat S_i$. The starting position of the next segment in $Y$ is $p_i+b$.
\end{enumerate}
\item The decoder now checks the last bit of the decoded segment $\hat s_{i,b}$. If $\hat s_{i,b} = 0$, the decoder knows that segment $(i+1)$ has been drawn from 
$\Ac^{1}$; otherwise it has been drawn from $\Ac^{0}$. Thus the decoder is now ready to decode segment $(i+1)$.

\end{enumerate}

\subsection{Non-binary Code Construction} 
We now construct segmented deletion correcting codes for alphabet size $q >2$.
For $a=0,\dotsc,b-1$, and  $c=0,\dotsc,q-1$, define following sets:
\be
\Ac^j_{a,c} \triangleq \big\{S \in \Xc^b : \syn(A_S)=a,\, \summ(S)=c,\, s_1,s_2\in \Xc\setminus\{j\} \big\}.
\label{eq:Ajac}
\ee
for $j=0,\dotsc,q-1$.
Now for each $j=0,\dotsc,q-1$ define
\be 
\{a_j,c_j\} = \argmax_{\substack{0\leq a \leq b-1 \\ 0\leq c \leq q-1}} |\Ac^j_{a,c}|.
\ee
Similarly to the binary case, the sets $\Ac^j_{a_j,c_j}$ for $0\leq j \leq q-1$ are used to construct the codebook. Choose the first segment from $A^0_{a_0,c_0}$. For encoding $i$th segment ($i>1$) we choose a word from $\Ac^j_{a_j,c_j}$ if $j$ is the last symbol of segment $i-1$. 
 The size each set  $\Ac^j_{a_j,c_j}$, for $0\leq j \leq q-1$, can be bounded from below as
\begin{align} 
M_s&\geq \frac{q^{b-2}(q-1)^2}{qb}.
\end{align}
Indeed, for any $j \in \{0,(q-1)\}$, there are $q^{b-2} (q-1)^2$ sequences of length $b$ with the first two symbols are not equal to $j$. Each of these symbols belong to one of the sets $\Ac^j_{a,c}$, where  $0\leq a \leq b-1$, and  $0 \leq c \leq q-1$. Therefore the largest set has size at least $\frac{q^{b-2}(q-1)^2}{qb}$.
This gives a lower bound on the rate
\be
R\geq \log_2 q-\frac{1}{b}\log_2 b-\frac{1}{b}\log_2 q -\frac{2}{b}\log_2\left(\frac{q}{q-1}\right).
\ee

Decoding proceeds in a  similar way to the binary case. The main difference is that instead of computing \eqref{eq:bin_dec_del}, the decoder computes
\be
\hat a= \syn(A_Z), \quad \hat c=\summ(Z)
\ee
where 
\be
Z=Y(p_i+1:p_i+b).
\ee
Then, the conditions in cases 1) a)  and 1) b) are replaced by $\{ \hat a = a_0 \text{ and } \hat c = c_0 \}$ and  by $\{ \hat a \neq a_0 \text{ or } \hat c \neq c_0\}$, respectively.

\section{Segmented Insertion Correcting Codes}\label{sec:ins}
\subsection{Binary Code Construction}
\label{subsec:binary_insertion}
As in the deletion case, we define  a  subset of VT codewords such that  upon decoding a segment, there is no ambiguity  in the starting position of the next segment. We define the following set of sequences
\begin{align}\label{ins-set}
\Ac_a \triangleq \bigl\{S\in \{0,1\}^b: \syn(S) = a,\,
 s_1s_2=01,\, s_3s_4\neq 01, \,S\neq 011\cdots 1 \bigr\}
\end{align}
and 
\be
a_0 = \argmax_{0\leq a\leq b} |\Ac_a|.
\ee
Similarly to the previous section, the sets $\Ac_a\subseteq\Vc\Tc_a(b)$ are sets of VT codewords with a prefix of a certain form. 
Our code is thus the maximal code in this family, i.e., $\Cc = \Ac_{a_0}^k$.  In contrast to the deletion case, the codeword for each segment is drawn from the same set $\Ac_{a_0}$.

In order to find the size of the code, we use similar arguments to those in the previous section. There are $2^{b-2}$ sequences with prefix $01$, out of which $2^{b-4}$ are removed because they have prefix $0101$; $01\cdots1$ is excluded from $\Ac_{a}$ by construction. Each of the $2^{b-2}-2^{b-4}-1$ sequences belong to exactly one of the sets $\Ac_{0},\dotsc,\Ac_{b}$. Therefore, the largest of these $b+1$ sets will have size at least
\begin{equation}
\label{eq:ins_size_bound}
\abs{\Ac_{a_0}} \geq \frac{2^{b-2}-2^{b-4}-1}{b+1}.
\end{equation}
This yields the following lower bound for the rate for  $b\geq 6$:
\begin{equation}
R\geq 1-\frac{1}{b}\log_2 (b+1)-\frac{2.5}{b}.
\label{eq:rate_ins}
\end{equation}
Hence the rate penalty is at most $\frac{2.5}{b}$ due to the added constraints on the prefix.

\subsection{Decoding}
Decoding proceeds on a segment-by-segment basis, and as in the case of deletions, the code structure ensures that before decoding segment $i$, the previous $(i-1)$ segments have been correctly decoded. Thus the decoder knows the correct starting position of segment $i$ in $Y$; as before, denote it by $p_i+1$.

\begin{enumerate}
\item The decoder computes the VT syndrome 
\be
\hat a = \syn\bigl(Y(p_i+1:p_i+b)\bigr)
\ee
and compares it to the correct syndrome $a_0$. There are two possibilities:

\begin{enumerate}
\item $\underline{\hat a \neq a_0}$: The decoder knows that there has been an insertion in this segment and feeds  $Y\bigl(p_i+1:p_i+b+1\bigr)$ to the VT decoder to recover the codeword. The output of the VT decoder is the decoded segment $\hat S_i$. The decoder proceeds decoding segment $i+1$, skipping step $2$. The starting position in $Y$ for decoding segment $i+1$ is $p_i +b +2$.

\item $\underline{\hat a = a_0}$: The decoder concludes that there is no insertion in $Y\bigl(p_i+1:p_i+b\bigr)$. This is because if there was an insertion in segment $i$, then $Y(p_i+1: p_i+b)$ cannot have VT syndrome $a_0$ unless $Y(p_i+1: p_i+b) = S_i$ --- indeed, if $Y(p_i+1: p_i+b) \neq S_i$, then both these length $b$ sequences would have syndrome $a_0$ and $Y(p_i+1: p_i+b+1)$ as a supersequence, which contradicts the property of VT codes in \eqref{eq:D1I1_prop}.

 In this case, the decoder outputs $\hat S_i = Y\bigl(p_i+1:p_i+b\bigr)$.  

\end{enumerate}
\item If case  1.b) holds, the decoder has to check whether $y_{p_i +b +1}$ could be an inserted bit at the very end of the segment. To this end, the $Y(p_i+b+1:p_i+b+4)$ is checked against the prefix conditions for segment $i+1$ set in $\Ac_{a_0}$. 
\begin{enumerate}
\item If $\underline{y_{p_i+b+1} y_{p_i+b+2}\neq 01}$: the decoder understands that there is an irregularity caused by either an insertion in $y_{p_i+b+1}$, or in $y_{p_i+b+2}$ or both. Therefore it deletes $y_{p_i+b+1}$ and proceeds to decode segment $i+1$ starting from  $y_{p_i+b+2}$. 
\item If $\underline{y_{p_i+b+1} y_{p_i+b+2} = 01,\,  y_{p_i+b+3}y_{p_i+b+4}\neq 01 }$, then $y_{p_i+b+1}$ is the correct start of segment $i+1$.
\item If $\underline{y_{p_i+b+1} y_{p_i+b+2} = 01,\,  y_{p_i+b+3}y_{p_i+b+4}= 01 }$: In this case, the decoder needs to decide among three alternatives by decoding segment $i+1$:
\begin{enumerate}
\item $y_{p_i+b+3}=0$ is an inserted bit in segment $i+1$ and no inserted bit in segment $i$; let $\tilde Y_1 = y_{p_i+b+1} y_{p_i+b+2} y_{p_i+b+4} \cdots y_{p_i+2b+1}$ denote the length $b$ sequence resulting from deleting $y_{p_i+b+3}$ from the received sequence. If $\syn(\tilde Y_1) = a_0$ then $\hat S_{i+1} = \tilde Y_1$. 
\item $y_{p_i+b+4}=1$ is an inserted bit in segment $i+1$ and no inserted bit in segment $i$;  let $\tilde Y_2 = y_{p_i+b+1} y_{p_i+b+2} y_{p_i+b+3} y_{p_i+b+5}\cdots y_{p_i+2b+1}$ denote the length $b$ sequence resulting from deleting $y_{p_i+b+4}$ from the received sequence. If $\syn(\tilde Y_2) = a_0$ then $\hat S_{i+1} = \tilde Y_2$. 
\item $y_{p_i+b+1}=0$, $y_{p_i+b+2}=1$ are inserted bits in segments $i$ and $i+1$, respectively; let $\tilde Y_3 = y_{p_i+b+3} y_{p_i+b+4}\cdots y_{p_i+2b+2}$ denote the length $b$ sequence resulting from deleting $y_{p_i+b+1}, y_{p_i+b+2}$ from the received sequence. If $\syn(\tilde Y_3) = a_0$ then $\hat S_{i+1} = \tilde Y_3$. 
\end{enumerate}

\end{enumerate}
\end{enumerate}

When $Y(bi+1:bi+4)= 0101$, we now show that the three cases listed in step 2.c) are mutually exclusive, and hence only one of them will give a matching VT syndrome. What needs to be checked is that the syndromes of $\tilde Y_1, \tilde Y_2, \tilde Y_3$ will all be different. From the very properties of VT codes we know that $\syn(\tilde Y_1)\neq \syn(\tilde Y_2)$. Now find that
\begin{align}
&\syn(\tilde Y_1) - \syn(\tilde Y_3) ~~~~(\text{mod}  (b+1))\\
& = \sum_{j=1}^b j \,\tilde y_{1,j}  - \sum_{j=1}^b j \,\tilde y_{3,j} ~~~~(\text{mod}  (b+1))\\
&= 5 + \sum_{j=p_i +b+5}^{p_i+2b+1} y_j - 2 - b y_{p_i+2b+2}~~~~~~(\text{mod}  (b+1))\\
& = 3 + w_H\bigl( Y (p_i+b+5:p_i+2b+1)\bigr) + y_{p_i+2b+2} ~~~~~~(\text{mod}  (b+1))\\
&\neq 0
\label{eq:insY1Y3}
\end{align}
where $w_H(Z)$ denotes the Hamming weight of sequence $Z$. The last step of \eqref{eq:insY1Y3} holds  because 
\be
 3 + w_H\bigl( Y (p_i+b+5:p_i+2b+1)\bigr) + y_{p_i+2b+2} \  (\text{mod}   (b+1)) 
 \ee
can equal to $0$ only if $w_H\bigl( Y (p_i+b+5:p_i+2b+1)\bigr) = b-3$ and $y_{p_i+2b+2}=1$, implying that both $\tilde Y_1=\tilde Y_3=011\cdots1$. Since this sequence has been explicitly excluded from the codebook, we always have strict inequality, and hence $\syn(\tilde Y_1) \neq \syn(\tilde Y_3)$. Furthermore, since
\be
\syn(\tilde Y_2) - \syn(\tilde Y_3) = \syn(\tilde Y_1) - \syn(\tilde Y_3) -1
\ee
is always non-zero, we conclude that there is no  ambiguity at the decoder .

\subsection{The Liu-Mitzenmacher conditions for binary segmented codes} \label{subsec:LM}

In \cite{liuMitz10}, Liu and Mitzenmacher specified three conditions such that any set of binary sequences satisfying these conditions is a zero-error code for both the segmented insertion channel and the segmented deletion channel. We list these conditions in Appendix \ref{app:LM}, and show that the segmented insertion correcting code $\Ac_{a_0}$ described in Sec. \ref{subsec:binary_insertion} satisfies these conditions. This shows that the segmented insertion correcting code can also be used for the segmented deletion channel, with the decoder proposed in  \cite{liuMitz10}.  The deletion correcting code described in Section \ref{sec:del} has a slightly higher rate than the insertion correcting code in in Sec. \ref{subsec:binary_insertion}. Moreover, the construction for the deletion case is more direct and can be generalized to non-binary  alphabets and the segmented  insertion-deletion channel.

However, the binary deletion correcting code proposed in Sec. \ref{subsec:binary_deletion} (or more precisely, the combined set of codewords $\Ac^0_a \cup \Ac^1_a$) cannot be guaranteed to satisfy the Liu-Mitzenmacher conditions. Therefore,  the construction in Sec. \ref{subsec:binary_deletion} may not be a zero-error code for the segmented insertion channel, even with an optimal decoder. 

It was conjectured in \cite{liuMitz10}  that the rate and size of the maximal code satisfying the three  sufficient conditions grows with $b$. As our insertion correcting code $\Ac_{a_0}$ satisfies the sufficient conditions, the lower bounds on its rate and size given in \eqref{eq:ins_size_bound} and \eqref{eq:rate_ins} confirm this conjecture.

\subsection{Non-binary Code Construction}
For the  segmented insertion channel with alphabet size $q>2$, we use prefix VT codes similar to those for the binary case. In this case, however, we set a prefix of length 3. This incurs a small penalty in rate with respect to the binary code described in Section \ref{subsec:binary_insertion}, but results in a slightly simpler decoder. Define the following sets for all $a=0,\dotsc,b-1$ and $c=0,\dotsc,q-1$
\begin{align}\label{ins-setq}
\Ac_{a,c} \triangleq \bigl\{S\in \Xc^b:  \syn(A_S)=a, \,\summ(S)=c, \,s_1s_2s_3=001 \bigr\}.
\end{align}
Now choose the largest set as the codebook, i.e., $\Cc = \Ac_{a_0,c_0}$ where
\be 
\{a_0,c_0\} = \argmax_{\substack{0\leq a \leq b-1 \\ 0\leq c \leq q-1}} |\Ac_{a,c}|.
\ee
Similar to the binary case, the number of codewords can be bounded from below as
\be
M_s\geq\frac{q^{b-3}}{qb},
\ee
which gives the following lower bound on the rate:
\begin{equation}
R\geq \log_2 q-\frac{1}{b}\log_2 b-\frac{4}{b}\log_2 q.
\end{equation}

Decoding proceeds in a similar manner to the binary case.  As the code is somewhat different from the binary one, we give a few more details about the decoder. Assume that the first $(i-1)$ segments have been decoded correctly, and  let $p_i+1$ is the starting point of the $i$th segment. Let 
\be
Z=Y(p_i+1:p_i+b),
\ee 
and compute
\be
\hat a= \syn(A_Z), \quad \hat c=\summ(Z).
\ee
\begin{enumerate}
\item $\underline{\hat a \neq a_0 \text{ or } \hat c \neq c_0} $: The decoder knows there has been an insertion in the $i$th segment and feeds $Y\bigl(p_i+1:p_i +b+1\bigr)$ to the non-binary VT decoder to recover the codeword. The output of the VT decoder is the decoded segment $\hat S_i$. The starting position of the next segment in $Y$ is $p_i+b+2$.
\item $\underline{\hat a = a_0 \text{ and } \hat c = c_0} $: The decoder concludes that there is no insertion in segment $i$ and outputs $\hat S_i=Y\bigl(p_i+1:p_i+b\bigr)$. The decoder must then investigate the possibility of an insertion at the very end of the $i$th segment in order to find the correct starting point of the next segment. This is done as follows. First, if the symbol $y_{p_i+b+1}$ is not equal to $0$, it is an insertion. The decoder deletes the inserted symbol, and the starting position for the next segment is 
$(p_i+b+2)$. Next, if  $y_{p_i+b+1}=0$ and  there is any symbol different from 0 or 1 in position $(p_i+b+2)$ or $(p_i+b+3)$, it is an inserted symbol thanks to the binary prefix. The decoder deletes the inserted symbol and sets the starting position of the next segment to $(p_i+b+1)$.  If neither of these cases hold, the decoder  follows Table \ref{tab:eq}. 

\end{enumerate}

\begin{table}[h!]
\centering
\caption{State of $y_{p_i+b+1}$ when $\hat a = a_0 \text{ and } \hat c = c_0$.\vspace{1mm}}
\begin{tabular}{ccc} 
\toprule
$Y(p_i+b+1:p_i+b+3)$ & State of $y_{p_i+b+1}=0$  & Starting point of next segment\\
\midrule   
$001$ & No action ($y_{p_i+b+1}$ is not an insertion) & $p_i+b+1$\\ 
$000$ & Delete the first zero  ($y_{p_i+b+1}$ is an insertion) & $p_i+b+2$\\
$010$ & Delete the $1$ ($y_{p_i+b+1}$ may or may not be inserted) & $p_i+b+1$\\
\bottomrule 
\end{tabular}
\label{tab:eq}
\end{table}

\section{Segmented Insertion-Deletion Correcting Codes}\label{sec:indel}

\subsection{Binary Code Construction} 
Since we now have both insertion and deletions,  the decoder must first identify the type of edit in a segment prior to correcting it.
 Define the following sets:
\begin{align}
\Ac^0_a&\triangleq \bigl\{S\in \{0,1\}^b : \syn(S)=a, \,s_1s_2s_3s_4s_5= 00111, \,s_{b-2}=s_{b-1}=s_b\bigr\}\\
\Ac^1_a&\triangleq \bigl\{S\in \{0,1\}^b : \syn(S)=a, \, s_1s_2s_3s_4s_5= 11000,\, s_{b-2}=s_{b-1}=s_b\bigr\}.
\end{align}
As in previous sections, these are subsets of VT codewords with certain constraints. In this case, in order to be able to identify the edit type, both prefix and suffix constraints have been added. Based on the above sets, we further define
\be
a_0 = \argmax_{0\leq a \leq b} |\Ac^0_a|, \quad a_1 = \argmax_{0\leq a \leq b} |\Ac^1_a|
\ee
and $M_s = \min\{|\Ac^0_{a_0}|,|\Ac^1_{a_1}|\}$. We construct the sets $\Ac^0,\Ac^1$ by choosing $M_s$ sequences from $\Ac^0_{a_0},\Ac^1_{a_1}$, respectively. Finally, the overall code of length $n=kb$ is constructed by choosing a codeword for each segment from either $\Ac^0$ or $\Ac^1$. 
The codeword for the first segment is chosen from $\Ac^0$. For $i \in \{2,\dotsc,k\}$, if the last bit of segment $(i-1)$ is 0, then the codeword for segment $i$ is drawn from $\Ac^1$ and otherwise from $\Ac^0$.

The size and rate are lower-bounded using the same arguments as in the previous sections. For $b \geq 7$, we obtain 
\be
M_s \geq \frac{2^{b-7}}{b+1}
\ee
which yields a rate lower bound  given by
\begin{equation}
R\geq 1-\frac{1}{b}\log(b+1)-\frac{7}{b}.
\label{eq:rate_indel}
\end{equation}
Due to the prefix and suffix constraints, our segmented insertion-deletion correcting codes have a rate penalty of at most $\frac{7}{b}$.


\subsection{Decoding}

As in the previous two cases, decoding proceeds segment-by segment. We ensure that before decoding segment $i$, the previous $(i-1)$  segments have all been correctly decoded.  Hence, the decoder knows the correct starting position  in $Y$ for segment $i$, which is denoted by $p_i+1$. The decoder also knows whether $S_i$ belongs to $\Ac^0$ or to $\Ac^1$. We discuss the case where $S_i\in \Ac^0$, so $\mathsf{syn}(S_i)=a_0$; the case where $S_i\in \Ac^1$ is similar, with the roles of the bits reversed.

The decoder computes the syndrome $\syn\big(Y(p_i+1:p_i+b)\big)$, and checks whether it equals $a_0$. There are two possibilities:
\begin{enumerate}
\item  $\syn\big(Y(p_i+1:p_i+b)\big)\neq a_0$: This means that there is an edit in this segment, we should identify the type of edit and correct it. We show that can be done without ambiguity by using the fact that three last bits of each segment (suffix) are the same, and considering prefix of the next segment. The decoder's decision for each combination of the three consecutive bits $(y_{p_i+b-1},y_{p_i+b}, y_{p_i+b+1})$ is listed in Table \ref{tab:neqa0}.  Once the type of edit is known, the decoder corrects the segment using the appropriate VT decoder.  We now justify the decisions listed in Table \ref{tab:neqa0}.
\begin{enumerate}
 \item If \un{$y_{p_i+b-1}=y_{p_i+b}=y_{p_i+b+1}$}:  The edit is an insertion. To see this, assume by contradiction that it was a deletion. Then at least one of $y_{p_i+b}$ and $y_{p_i+b+1}$ are the first bit of the prefix of $S_{i+1}$, and  $y_{p_i+b-1}$  is a suffix bit  of $S_i$. This is not possible because by construction, the first two prefix bits of $S_{i+1}$ must be different from the suffix bits of $S_i$.
 
\begin{table*}[t!]
\centering
\caption{Type of edit when $\mathsf{syn}\big(Y(p_i+1:p_i+b)\big)\neq a_0$ \vspace{1mm}}
\begin{tabular}{cc} 
\toprule
State of sequence & Type of edit \\
\midrule   
$y_{p_i+b-1}=y_{p_i+b}=y_{p_i+b+1}$ &  Insertion \\
$y_{p_i+b-1}=y_{p_i+b}\neq y_{p_i+b+1}$ & Deletion \\
$y_{p_i+b-1}=y_{p_i+b+1}\neq y_{p_i+b}$ and $\mathsf{syn}(Z)\neq a_0$, where $Z=[Y(p_i+1:p_i+b-1),y_{p_i+b+1}]$ & Deletion  \\
$y_{p_i+b-1}=y_{p_i+b+1}\neq y_{p_i+b}$ and $\mathsf{syn}(Z)=a_0$ and $y_{p_i+b+1}=y_{p_i+b+2}=y_{p_i+b+3}$& Deletion  \\
$y_{p_i+b-1}=y_{p_i+b+1}\neq y_{p_i+b}$ and $\mathsf{syn}(Z)=a_0$ and ($y_{p_i+b+1}\neq y_{p_i+b+2}$ or $y_{p_i+b+1}\neq y_{p_i+b+3}$) & Insertion  \\
$y_{p_i+b-1}\neq y_{p_i+b}= y_{p_i+b+1}$ and $y_{p_i+b-2}=y_{p_i+b-1}$ & Deletion    \\
$y_{p_i+b-1}\neq y_{p_i+b}= y_{p_i+b+1}$ and $y_{p_i+b-2}\neq y_{p_i+b-1}$ &  Insertion     \\
\bottomrule 
\end{tabular}
\label{tab:neqa0}
\end{table*}

 \item If  \un{$y_{p_i+b-1}=y_{p_i+b}\neq y_{p_i+b+1}$}: The edit is a deletion. To see this, suppose that  the edit was an insertion; then the suffix condition can only be satisfied if  $y_{p_i+b+1}$ is the inserted bit. However, this implies that $\syn\big(Y(p_i+1:p_i+b)\big)= a_0$, which is contradiction.
 
 \item If \un{$y_{p_i+b-1}=y_{p_i+b+1}\neq y_{p_i+b}$}: The edit could be either an insertion, or a deletion, according to the rules  in lines 3, 4, 5 of Table \ref{tab:neqa0}.  If the the edit is an insertion, then $y_{p_i+b}$ is the inserted bit, therefore by omitting this bit, the sequence $Z=[Y(p_i+1:p_i+b-1),y_{p_i+b+1}]$ should have VT-syndrome equal to $a_0$. Therefore, if $\mathsf{syn}(Z) \neq a_0$, then the edit is deletion; if $\mathsf{syn}(Z) = a_0$, we need  to check the prefix of the next segment to determine the type of edit. 
 
 If \un{$\mathsf{syn}(Z)=a_0$}:  If $y_{p_i+b+1}=y_{p_i+b+2}=y_{p_i+b+3}$, then the edit in segment $i$ is a deletion (it can be verified that the prefix condition for segment $(i+1)$ cannot otherwise be satisfied with at most one edit),. In all other cases the edit in segment $i$ is insertion, with $y_{p_i+b}$ being the inserted bit. We observe that when $\mathsf{syn}(Z)=a_0$, $S_i=Z$ with either type of edit, but the decoder needs to  infer the type of edit in order to guarantee the correct starting position for the next segment.

\item If \un{$y_{p_i+b-1}\neq y_{p_i+b}= y_{p_i+b+1}$}: In this case, $y_{p_i+b-2}$ determines the type of edit: if $y_{p_i+b-2}=y_{p_i+b-1}$ the edit is a deletion, otherwise it is an insertion. This can be seen by examining the suffix condition: if the edit is an insertion then $y_{p_i+b-1}$ is the inserted bit therefore $y_{p_i+b-2}$ belongs to suffix of $S_i$, hence $y_{p_i+b-2}=y_{p_i+b}=y_{p_i+b+1}$. On the other hand, if the edit is a deletion, then $y_{p_i+b-2}$ and $y_{p_i+b-1}$ belongs to suffix of $S_i$, so they should be equal.
 \end{enumerate}
 
\begin{table}[t!]
\centering
\caption{State of $y_{p_i+b+1}$ when $\mathsf{syn}(Y(p_i+1:p_i+b)) = a_0$.\vspace{1mm}}
\begin{tabular}{cc} 
\toprule
$Y(p_i+b+1:p_i+b+5)$ & State of $y_{p_i+b+1}$\\
\midrule   
$1uvst$ & Inserted \\
$000uv$ & Inserted \\
$011uv$ & Not Inserted \\
$01000$ & Not possible \\
$01001$ & Inserted \\
$01010$ & Not possible \\
$01011$ & Not inserted \\
$00100$ & Not possible \\
$00101$ and $\mathsf{syn}(Z_1)$ matches &Inserted \\
$00101$ and $\mathsf{syn}(Z_2)$ matches &Not Inserted \\
$00110$ & Not Inserted \\
$00111$ & Not Inserted \\
\bottomrule 
\end{tabular}
\label{tab:equala0}
\end{table}

Hence we have shown that whenever $\syn\big(Y(p_i+1:p_i+b)\big)\neq a_0$, we can uniquely decode $S_i$ and determine the correcting starting position  for the next segment. 

\item $\mathsf{syn}\big(Y(p_i+1:p_i+b)\big)= a_0$: In this case, by combining the arguments in step 1.a) of  the deletion decoder and step 1.b) of the insertion encoder, we conclude that $\hat S_i=Y(p_i+1:p_i+b)$. To determine the correct starting position for the next segment, we have to investigate the possibility of an insertion at the end of the block, i.e., determine whether $y_{p_i+b+1}$ is an inserted bit. This can be done by examining the prefix of $S_{i+1}$. We consider five bits, $Y(p_i+b+1:p_i+b+5)$, and for all 32 cases determine the state of $y_{p_i+b+1}$. For the simplicity, assume that  the last bit of $S_i$ is $1$,  so that  the prefix for $S_{i+1}$ is $00111$; the other case is identical, with $0$
and $1$ interchanged.

First, if $y_{p_i+b+1} = 1$, then it is  an inserted bit (this is 16 of the 32 cases).  Table \ref{tab:equala0} lists the type of edit for each of the other cases when $y_{p_i+b+1} = 0$. These are justified below.
\begin{enumerate}
\item If \un{$Y(p_i+b+1:p_i+b+5) =011uv$} for some bits $u,v$, then $y_{p_i+b+1}$ is not an insertion corresponding to segment $i$: if it was inserted, then decoding for  segment $(i+1)$ would start with the bits $11\ldots$, which cannot be matched with the prefix $00111$ with only one edit.  Hence the correct starting position for decoding segment $(i+1)$ is $p_{i} +b +1$.

\item If \un{$Y(p_i+b+1:p_i+b+5) = 000uv$}, then $y_{p_i+b+1}$ (or another $0$ from the run) is an insertion for segment $i$, as $000u$ does not match $0011$ unless  we remove a zero form the run.

\item The cases \un{$Y(p_i+b+1:p_i+b+5) = 01000$,$01010$}, \un{$00100$} cannot occur as they cannot be matched with the required prefix $00111$ through any valid edits for segment $i+1$,  whether or not $y_{p_i+b+1}$ is inserted.

\item If \un{$Y(p_i+b+1:p_i+b+5) = 01001$ },  then  $y_{p_i+b+1} $ is an insertion for segment $i$ as this is the only option consistent with the prefix $00111$. 

\item If \un{$Y(p_i+b+1:p_i+b+5) = 0011u$ or $01011$}, then   $y_{p_i+b+1}=0$ is not an insertion for segment $i$, and is the starting bit for decoding segment $(i+1)$. 

\item  If \un{$Y(p_i+b+1:p_i+b+5) =00101$}, we need to compare the VT syndromes of two sequences to determine the status of $y_{p_i+b+1}$. We will also decode $S_{i+1}$ in the process. If $y_{p_i+b+1}=0$ is inserted, then $y_{p_i+b+3}=1$ should also be inserted, therefore $S_{i+1}=Z_1$ where $Z_1=[00,Y(p_i+b+5:p_i+2b+2)]$.  On the other hand, if  $y_{p_i+b+1}$ is not inserted then $y_{p_i+b+4}=0$ should be an inserted bit, therefore, $S_{i+1}=Z_2$ where $Z_2=[001,Y(p_i+b+5:p_i+2b+1)]$.  However, $Z_1$ and $Z_2$ will always produce different syndromes and only one of them will be equal to $a_0$, the correct syndrome for segment $(i+1)$. Thus we can correctly identify whether $y_{p_i+b+1}$ was an insertion for segment $i$ or not.
\end{enumerate}
Hence we have shown that whenever $\mathsf{syn}(Y(p_i+1:p_i+b)) = a_0$, we can uniquely decode $S_i$ and determine the correcting starting position for the next segment.
\end{enumerate}

The decoding algorithm described above was  simulated in Matlab to confirm that the code is indeed zero-error. The Matlab files for implementing the  codes proposed for all three binary segmented edit models  are available at \cite{SegSimulator}.

\subsection{Non-binary Code Construction}
We now construct segmented  insertion-deletion correcting codes for  alphabet size $q >2$.
For $a=0,\dotsc,b-1$, and  $c=0,\dotsc,q-1$, define following sets:
\begin{align}
\Ac^0_{a,c} \triangleq \big\{S \in \Xc^b : \syn(A_S)=a, \,\summ(S)=c, \,s_1s_2s_3s_4s_5=00111, \,s_{b-2}=s_{b-1}=s_b \big\}, \\
\Ac^1_{a,c} \triangleq \big\{S \in \Xc^b : \syn(A_S)=a, \,\summ(S)=c, \,s_1s_2s_3s_4s_5=11000, \,s_{b-2}=s_{b-1}=s_b\big\}.
\end{align}
For $j=0,1$ define
\be 
\{a_j,c_j\} = \argmax_{\substack{0\leq a \leq b-1 \\ 0\leq c \leq q-1}} |\Ac^j_{a,c}|.
\ee
We use the sets $\Ac^0_{a_0,c_0}$ and $\Ac^1_{a_1,c_1}$ to construct the codebook by alternating depending on the last symbol of the previous segment. We set $M_s = \min\{\Ac^0_{a_0,c_0},\Ac^1_{a_1,c_1}\} $ and construct the sets $\Ac^0,\Ac^1$ by choosing $M_s$ sequences from $\Ac^0_{a_0,c_0},\Ac^1_{a_1,c_1}$, respectively. 
The codeword for the first segment is chosen from $\Ac^0$. For $i \in \{2,\dotsc,k\}$, if the last symbol of segment $(i-1)$ is an \emph{even} number, then the codeword for segment $i$ is drawn from $\Ac^1$; if the last symbol of segment $(i-1)$ is an \emph{odd} number, the codeword is drawn from $\Ac^0$.

The number of codewords per segment satisfies
\be
M_s\geq\frac{q^{b-7}}{qb}
\ee
and thus a lower bound on the rate is
\be
R\geq \log_2q-\frac{1}{b}\log_2 b-\frac{8}{b}\log_2 q.
\ee

The decoding is almost identical to the binary case. As with previous decoders, to decode segment $i$, it is assumed that the first $(i-1)$ segments have been decoded correctly.  Let $Z=Y(p_i+1:p_i+b)$,
where $p_i+1$ is  the starting position of the $i$th segment.
Compute
\be
\hat a= \syn(A_Z), \quad \hat c=\summ(Z).
\ee
The decoder checks whether $\{\hat a = a_0 \text{ and } \hat c = c_0 \}$ or $\{ \hat a \neq a_0 \text{ or } \hat c \neq c_0 \}$. In the first case, the decoder sets $\hat S_i=Y(p_i+1:p_i+b)$ and in order to find the starting point of segment $i+1$, follows the same case breakdown as in the binary decoder (see case 2 of the binary decoder). On the other hand, if $\{ \hat a \neq a_0 \text{ or } \hat c \neq c_0 \}$, thanks to the prefix-suffix code structure being the same as the binary one, the decoder follows exactly the same case breakdown (see case 1 of the binary decoder) in order to identify the type of edit and correct it.

\section{Conclusion}

We have considered three segmented edit channel models and proposed zero-error codes for each of them over alphabets of size $q \geq 2$. The proposed codes are constructed using carefully chosen subsets of VT codes, and can be decoded in a segment-by-segment fashion in linear time. The rate scaling for the  codes is shown to be the same as that of the maximal code; the upper bound of Theorem \ref{thm:rate_UB} shows that  the rate penalty is of order $1/b$. 

One direction for future work is to obtain tighter non-asymptotic upper and lower bounds on the cardinality of these codes. For tighter upper bounds, the linear programming technique from \cite{kulkarniK13} is a promising approach.  For tighter lower bounds, one approach would be to use the known formulas for the cardinality of VT codes \cite{Sloane00}, and adapt them to our setting where prefix and/or suffix constraints are added.

\appendix

\subsection{Proof of \eqref{eq:binom_tail}} \label{app:binom_tail} 

Let $U$ be a $\text{Binomial}\left( b, \frac{q-1}{q} \right)$  random variable, with mean $\mu = \frac{b(q-1)}{q}$. Then, using a standard Chernoff bound for a binomial random variable (see, for example \cite[Theorem 4.5]{mitzbook}), we have for any $\e>0$:
\be
\mathbb{P}(U  \leq \mu(1- \e) ) \leq \exp\left(   \frac{- \mu\e^2}{2}  \right).
\label{eq:U_chernoff}
\ee
Choosing $\e =  \sqrt{ \frac{2  \kappa q \ln b}{(q-1)b}}$, we have
\begin{align}
\mu(1-\e) &=  \frac{b(q-1)}{q} -   \sqrt{ \frac{2  \kappa (q-1)b \ln b}{q}}\\
& =r,
\end{align} 
where $r$ is defined in \eqref{eq:rdef}. Using this in \eqref{eq:U_chernoff}, we obtain 
\begin{align}  
\mathbb{P}(U  \leq r)  &= \mathbb{P}(U  \leq \mu(1- \e))  \\
&\leq \exp\left(   \frac{- \mu\e^2}{2}  \right) \\
&= b^{-\kappa},\label{eq:PUlr0}
\end{align}
where the last equality is obtained by substituting the values of $\mu$ and $\e$.  Finally, note that
\be
\begin{split}
\mathbb{P}(U  \leq r)  &  \geq \mathbb{P}(U  \leq (r-1)) \\
& = \sum_{t=0}^{r-1} \left(1-\frac{1}{q}\right)^{t} \left( \frac{1}{q} \right)^{b-t} {b \choose t}  \\
& \geq \sum_{t=0}^{r-1} \left(1-\frac{1}{q}\right)^{t} \left( \frac{1}{q} \right)^{b-t} {b-1 \choose t}.
\end{split}
\label{eq:PUlr1}
\ee
Combining \eqref{eq:PUlr1} and \eqref{eq:PUlr0} yields the desired inequality.


\subsection{The Liu-Mitzenmacher conditions} \label{app:LM}

 Let $\Ic_1(X)$ denote the set of all sequences obtained by adding one bit to the binary sequence $X$.  Then $\Cc \subseteq \{0,1 \}^b$ is a binary zero-error code for both the segmented insertion channel and the segmented deletion channel (with segment length $b$) if the following conditions are satisfied.
\begin{enumerate}
\item For any $U,V\in \Cc$, with $U\neq V$, $\Ic_1(U)\cap \Ic_1(V)=\emptyset$;
\item For any $U,V\in \Cc$, with $U\neq V$, prefix$(\Ic_1(U))\cap$ suffix$(\Ic_1(V))=\emptyset$;
\item Any string of the form $y^*(zy)^*$ or $y^*(zy)^*z$, where $y,z \in\{0,1\}$, is not in $\Cc$.
\end{enumerate}
Here prefix$(X)$ denotes the subsequence  of $X$  obtained excluding the last bit, suffix$(X)$ the subsequence obtained excluding the first bit, and $X^*$ is the regular expression notation referring to $0$ or more copies of sequence $X$. The set prefix$(\Ic_1(U))$ is defined as $\{ \text{prefix}(X): X \in  \Ic_1(U) \}$. The set suffix$(\Ic_1(V))$ is defined similarly.

We now show that the insertion correcting code $\Ac_{a_0}$ defined in Sec. \ref{subsec:binary_insertion} satisfies these conditions. 
Since $\Ac_{a_0}$ is a subset of a VT code and is hence a single insertion correcting code, the first condition is satisfied.  

We next verify the third condition. All the codewords in $\Ac_{a_0}$ start with $01$. It is easy to see that any sequence starting with  $01$ and  violating  the third condition in either of the two ways must have $0101$ as its first four bits. But these sequences are excluded from $\Ac_{a_0}$, so each codeword in $\Ac_{a_0}$  satisfies the third condition.

It remains to prove that the second condition is satisfied. 
Assume towards contradiction that there exist  codewords $U,V \in \Ac_{a_0}$ such that $U \neq V$ and the set $\mathcal W = \text{ prefix}(\Ic_1(U)) \cap \text{suffix}(\Ic_1(V))$ is non-empty. Suppose that the sequence $Z \in \mathcal W$, and $Z_1\in \Ic_1(U)$ and $Z_2\in\Ic_1(V)$ are length $(b+1)$ sequences such that   that $Z=\text{prefix}(Z_1)=\text{suffix}(Z_2)$.

Since $U \in \Ac_{a_0}$ and $Z_1\in \Ic_1(U)$, $\text{prefix}(Z_1)$ will start with a $0$, unless the inserted bit in $Z_1$ is a $1$ and is inserted exactly at the beginning of $U$, i.e., unless $Z_1=[1,U]$. Also, since $Z_2\in\Ic_1(V)$, $\text{suffix}(Z_2)$ will start with $1$ unless $Z_2$ is obtained by adding a bit at the beginning of $V$, i.e. $Z_2=[h,V]$, for $h \in \{0,1\}$. Since $Z=\text{prefix}(Z_1)=\text{suffix}(Z_2)$, clearly one of the above two cases should hold. 
First, assume that $Z$ starts with $1$ and therefore we have $Z_1=[1,U]$. Now since $U\in \Ac_{a_0}$ starts with $01$, we have 
\begin{align}
Z&=\text{prefix}(Z_1)\\
&=Z_1(1:b)\\
&=[1,U(1:b-1)]\\
&=[101,U(3:b-1)].
\end{align}
Now we also know that $Z=\text{suffix}(Z_2)$, so $\text{suffix}(Z_2)=[101,U(3:b-1)]$. Now, notice that $Z_2\in\Ic_1(V)$ and first bit of $V$ is $0$, so the first two bits of $Z_2$ cannot be $11$. We therefore have
\be
Z_2=[0101,U(3:b-1)].
\ee
But we know that $V \in \Ac_{a_0}$ cannot start with $0101$, so either the third or the fourth bit in $Z_2$ is the inserted bit. Therefore, we know that 
\be
V=[01z,U(3:b-1)],
\ee
for $z\in \{0,1 \}$. We also know that
\be
U=[01,U(3:b-1),u_b],
\ee
where $u_b \in \{0,1\}$. But this contradicts condition 1 (which has already been verified) because we obtain the same length $(b+1)$ sequence by: i) inserting $u_b$ to the end of $V$, and ii) inserting $z$ after the second bit of $U$.


Next consider the second case where $Z$ starts with a $0$. As explained above, we then have $Z_2=[h,V]$, and hence, $Z= \text{suffix}(Z_2)=V$. Therefore $\text{prefix}(Z_1)=V$, so one can obtain $Z_1$ by adding the last bit of $Z_1$ to $V$. Therefore $Z_1\in\Ic_1(U)\cap \Ic_1(V)$, which is a contradiction. This completes the proof that 
$\Ac_{a_0}$ satisfies all the three conditions. 

\section*{Acknowledgement}

The authors thank the associate editor and the two anonymous referees for several helpful comments which led to an improved paper.

\bibliographystyle{ieeetr}
\bibliography{indel}
\end{document}